\documentclass{PoS}
\usepackage{amsmath}

\title{Monte Carlo methods in continuous time for lattice Hamiltonians}

\ShortTitle{Monte Carlo methods in continuous time for lattice Hamiltonians}

\author{\speaker{Emilie Huffman}\thanks{In collaboration with Shailesh Chandrasekharan.}\\
        Duke University\\
        E-mail: \email{emilie.huffman@duke.edu}}

\abstract{We solve a variety of sign problems for models in lattice field theory using the Hamiltonian formulation, including Yukawa models and simple lattice gauge theories. The solutions emerge naturally in continuous time and use the dual representation for the bosonic fields. These solutions allow us to construct quantum Monte Carlo methods for these problems. The methods could provide an alternative approach to understanding non-perturbative dynamics of some lattice field theories.}

\FullConference{34th annual International Symposium on Lattice Field Theory\\
		24-30 July 2016\\
		University of Southampton, UK}

\begin{document}

\section{Introduction}
Many-body quantum systems remain an area with much room for exploration. While quantum Monte Carlo (QMC) can be very effective for some models and sets of parameters, in most cases the sign problem, which causes the calculation time to scale exponentially with system size, remains a formidable barrier to computation. In addition, a general solution to sign problems is NP-hard. However, solving the sign problem for specific models is not always NP-hard and some good progress has been made recently in terms of solving sign problems in interacting fermionic systems. Solutions to the sign problem in the $t$-$V$ model were found in \cite{Huf14,Li:2014tla}, and since then much progress has been made in characterizing the mathematical structures that lead to solvability. \cite{Wang25,Wei:2016sgb, Li:2016gte}

Here we move beyond purely fermionic models to see how the above class of solvable models may be extended to include interacting hardcore bosons as well. We examine a variety of lattice models, including interacting spin and fermion models as well as lattice gauge theories, in the Hamiltonian approach. Condensed matter problems are more naturally formulated in this approach, which has certain advantages that include the ability to compute in continuous time. From a lattice field theory perspective, computing in continuous time has a potential advantage in that it results in reduced fermion doubling (because no doubling occurs in the time dimension). However, we note that discrete time calculations may be performed in all of these Hamiltonian approach solutions as well.

Much of the recent progress in the expansion of sign problem solutions through the Hamiltonian formalism has been made either through auxiliary field methods (CT-AUX and LCT-AUX), or interaction expansions (CT-INT and LCT-INT). Our new solutions arise in the CT-INT formalism, which consist of writing the full Hamiltonian $H = H_0 + H_{\rm int}$, where $H_0$ is designated the free part, and $H_{\rm int}$ is designated the interaction part, and then expanding the partition function as
\begin{equation}
Z = \sum_k \int \left[dt\right] (-1)^k {\rm Tr}\left(e^{-\left(\beta - t_1\right) H_0} H_{\rm int} e^{-\left(t_1 - t_2 \right) H_0} H_{\rm int}...\right).
\label{zint}
\end{equation}
While a straightforward application of CT-INT scales as $\beta^3 N^3$, where $\beta$ is the extent of the imaginary time
and N is the number of spatial lattice sites, it is possible to recast it in a more efficient form known as LCT-INT that scales as $\beta N^3$.\cite{Wang:2015rga} The essential idea behind why our models are solvable in CT-INT was originally introduced in lattice field theory to solve a sign problem in a Yukawa model involving interacting fermions and bosons.\cite{Chandrasekharan:2012fk}
\section{Solvable Spin-Fermion Interaction Models}
Many of these spin-fermion interaction models which are newly solvable are extensions of the $t$-$V$ model that we solved recently \cite{Huf14}. We discuss a wide range of these spin-fermion interaction models from a condensed matter perspective in \cite{Huffman:2016nrx}. All of the models below assume a bipartite lattice.
\subsection{Ising Spins with Fermionic Interaction}
The first model involves a simple Ising spin interaction:
\begin{equation}
\begin{aligned}
H =& -t\sum_{\left\langle xy\right\rangle} \left(c_x^\dagger c_y + c_y^\dagger c_x\right) + V \sum_{\left\langle xy\right\rangle} \left(n_x-\frac{1}{2}\right) \left(n_y - \frac{1}{2}\right)\\
& +J\sum_{\left\langle xy\right\rangle} S^3_x S^3_y - \sum_x h_x \left(n_x - \frac{1}{2}\right) S_x^1.
\end{aligned}
\end{equation}
Even with this relatively simple model we can see the potential for some interesting physics, however. The first two terms, $H_0^f$ and $H_{\rm int}^f$, together form the $t$-$V$ model, which has been shown to be solvable in both CT-INT\cite{Huf14} and CT-AUX\cite{Li:2014tla}. At the critical point, there is a transition from a semimetal to an insulator and the critical exponents have been calculated using multiple methods. \cite{Wang:2014cbw, Li:2014aoa} The insulator phase has a charge density wave ordering, favoring fermionic occupation either on odd sites alone, or on even sites alone.

Moving on to the spin degrees of freedom, we see that the Ising interaction term $H_0^b$ favors an alignment of the spins along the $z$-axis. However, the final term in the Hamiltonian, $H_{\rm int}^{fb}$ will complicate things. This term involves an interaction between both the fermionic and spin degrees of freedom, and a charge density wave ordering among the fermions would cause this term to favor spins aligning in the $x$-direction. Assuming that $V_c$ is the critical coupling, we thus propose that while $\left\langle S^3\right\rangle$ would be consistently nonzero when $V < V_c$ (which would be a semimetal phase for the $t$-$V$ model), it would approach zero as $h_x$ increased for $V > V_c$, meaning a destroyed Ising order.

Now that we have discussed the physics, we proceed to show how the model is sign-problem-free in CT-INT. The partition function, given by \ref{zint}, will consist of terms in the following form:
\begin{equation}
(-1)^k {\rm Tr}\left(e^{-\left(\beta - t_1\right) H_0} H_{\rm int} e^{-\left(t_1 - t_2\right) H_0} H_{\rm int} ...H_{\rm int} e^{-t_k H_0}\right),
\end{equation}
where $H_0 = H_0^f + H_0^b$ and $H_{\rm int} = H^f_{\rm int} + H^{fb}_{\rm int}$. No sign problem occurs in this expansion, and a straightforward way to see this is to apply the following unitary transformation to the Hamiltonian,
\begin{equation}
H \rightarrow U^\dagger H U, \qquad U = \prod_x e^{i\left(1-\eta_x\right) S_x^3 \pi /2}.
\label{transf}
\end{equation}
Here $\eta_x$ is a parity factor, and is $+1$ on even sites and $-1$ on the odd sites in a bipartite lattice. All the terms remain unchanged except for the spin-fermion coupling, $H^{fb}_{\rm int}$, which is transformed into
\begin{equation}
H^{fb, U}_{\rm int} = \sum_x \eta_x \left(n_x - \frac{1}{2}\right) S_x^1.
\label{trans}
\end{equation}
Note that the fermionic part of this term is in the form of a staggered chemical potential, which would introduce no sign problem to the $t$-$V$ model.\cite{Huf14} In the CT-INT expansion, it is in fact possible to separate out expansion terms into a purely fermionic sector and purely spin sector. As an example, consider one of the terms in the expansion at order $k=2$ with two insertions of interactions: one insertion of $H^f_{\rm int}$ at $t_1$ and another insertion of $H_{\rm int}^{fb}$ at $t_2$:
\begin{equation}
\begin{aligned}
(-1)^2 {\rm Tr_b}&\left(e^{-\left(\beta -t_2\right) H_0^b } h_z S_z^1 e^{-t_2 H_0^b}\right)\\
& \times {\rm Tr_f}\left(e^{-\left(\beta - t_1\right) H_0^f} V\left(n_x - \frac{1}{2}\right)\left(n_y - \frac{1}{2}\right) e^{-\left(t_1 - t_2\right) H_0^f} \eta_z \left(n_z - \frac{1}{2}\right) e^{-t_2 H_0^f}\right).
\end{aligned}
\end{equation}
Because the spin operators commute with the fermionic operators, factorization into a purely fermionic trace and a purely spin trace is possible. The factorized partition function is:
\begin{equation}
Z = \sum_{k,\{z\},m,\left\{b\right\}}  \int \left[dt\right] G_b\left[k,\left\{z\right\}\right] G_f \left[k, \left\{z\right\},m,\left\{b\right\}\right],
\label{ctintf}
\end{equation}
where
\begin{equation}
G_b\left[k,\left\{z\right\}\right] = (-1)^k {\rm Tr_b}\left(e^{-\left(\beta - t_1\right)H_0^b}  h_{z_1} S_{z_1}^1 \times e^{-\left(t_1 - t_2\right) H_0^b} h_{z_2} S_{z_2}^1 .... h_{z_k} S_{z_k}^1 e^{-\left(t_{k}\right) H_0^b}\right),
\end{equation}
is the trace over the spin space, and depends on $k$ insertions of the interaction terms $h_z S_z^1$ at the times $t_1,t_2,...,t_l$. Similarly, 
\begin{equation}
G_f \left[k, \left\{z\right\}, m,\left\{b\right\}\right] = (-1)^m {\rm Tr_f}\left(...\eta_{k_1}\left(n_{k_1} -1 / 2\right) ...H_{\rm int}^f(b_1) ... H_{\rm int}^f(b_m) ...\eta_{z_k}\left(n_{z_k} -1 / 2\right) ...\right)
\label{ftrace}
\end{equation}
is the trace in the fermionic space and depends on $m$ insertions of the interaction bonds $H_{\rm int}^f(b\equiv \langle xy\rangle) = V\left(n_x - \frac{1}{2}\right) \left(n_y - \frac{1}{2}\right)$ and $k$ insertions of $\eta_z\left(n_z -\frac{1}{2}\right)$ from the fermion-spin interactions. One such configuration of insertions is labeled by $[k,\{z\},m,\{b\}]$. The presence of the free propagators $\mathrm{e}^{-t H_0^f}$ between these insertions are hidden in the ellipses. Note that for every insertion at of $S_z^1$ at $t_z$ in the spin space, we have a corresponding insertion of $\eta_z\left(n_z - \frac{1}{2}\right)$ at $t_z$ in the fermionic space. The partition function is thus a sum over all possible configurations $[k,\{z\},m,\{b\}]$.

\begin{figure}       
    \makebox{\includegraphics[width=6cm]{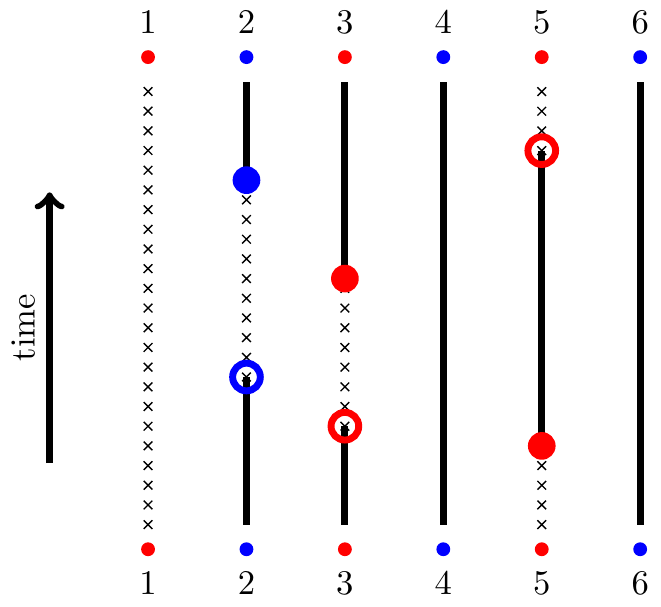}}   
    \hspace{40px}
    \makebox{\includegraphics[width=6cm]{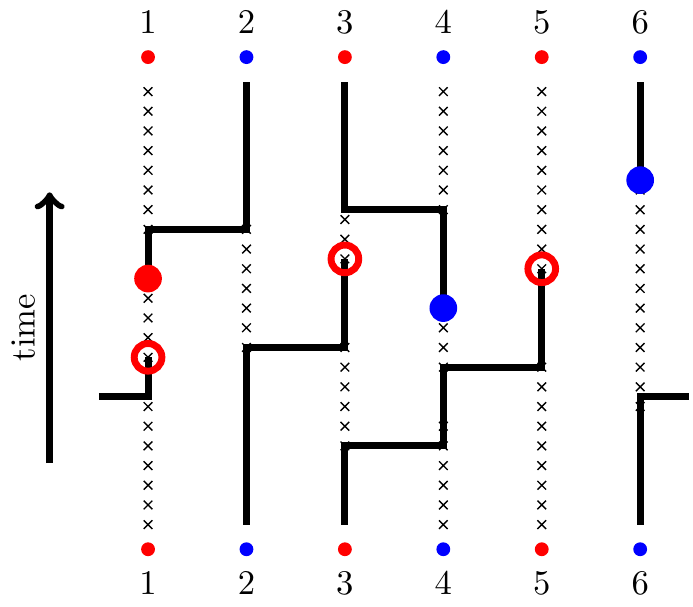}}
    \caption{A bosonic worldline diagram for the Ising model to the left, and one for the Heisenberg model on the right. The numbers label spatial sites. Insertions of $S^1$ create or annihilate hardcore bosons. Filled circles indicate creation events while the open circles indicate annihilation events. The number of $S^1$ operators is even due to temporal periodicity of the worldlines.}
    \label{worldlines}
\end{figure}

As alluded to earlier, the fermionic parts of the trace, $G_f \left[k, \left\{z\right\}, m,\left\{b\right\}\right]$, are the same as those for a $t$-$V$ model with a staggered potential,so they are all positive. As for the spin parts, we consider the $G_b\left[k,\left\{z\right\}\right]$ terms as traces over the spin $z$ basis. When we do that, we get
\begin{equation}
\begin{aligned}
& G_b\left[k,\left\{z\right\}\right] = (-1)^k \sum_{\left\{s^3\left(t\right)\right\}}  \left\langle s^3\left(t_0\right)\right| e^{-\left(\beta-t_1\right) H_0^b} h_{z_1} S_{z_1}^1 \left|s^3\left(t_1\right)\right\rangle  \\
\times & \left\langle s^3\left(t_1\right)\right| e^{-\left(t_1 -t_2\right) H_0^b} h_{z_2} S_{z_2}^1 \left|s^3\left(t_2\right)\right\rangle ... \left\langle s^3\left(t_k\right)\right| e^{-t_k H_0^b} \left|s^3\left(t_0\right)\right\rangle .
\end{aligned}
\end{equation}
where the sum over $\left\{s^3\left(t\right)\right\}$ indicates a sum over all space-time spin configurations that are periodic  i.e., $s^3\left(t_0\right) = s^3\left(t_k\right)$. The $S^1_z = \frac{1}{2} \left(S_z^- + S_z^+\right)$ operators serve to flip the spins between spin up and spin down states, thus periodicity ensures that the $h_z S_z^1$ insertions must come in pairs. Thus $k$ is even and the spin factors $G_b\left[k,\left\{z\right\}\right]$ are also guaranteed to be positive. Another way to understand the positivity is to consider quantum spins as hardcore bosons (with spin-up representing particles and spin-down representing their absence). Then for every creation of a particle caused by the $S_z^1$ operator, we require a corresponding annihilation of the same particle caused by a second $S_z^1$ operator to preserve the trace. The left side of figure \ref{worldlines} illustrates this. Because both spin and fermion traces can be evaluated in polynomial time, we conclude this model has no sign problem in this CT-INT formulation.
\subsection{Heisenberg Antiferromagnetic Spins with Fermionic Interaction}
The second model we consider has a somewhat more complicated spin section, consisting of a full $SU(2)$ symmetric antiferromagnetic interaction:
\begin{equation}
\begin{aligned}
H =& -t\sum_{\left\langle xy\right\rangle} \left(c_x^\dagger c_y + c_y^\dagger c_x \right) + V \sum_{\left\langle xy\right\rangle} \left(n_x - \frac{1}{2} \right) \left(n_y - \frac{1}{2}\right)\\
& + J\sum_{\left\langle xy\right\rangle}\vec{S}_x \cdot \vec{S}_y - \sum_x h_x \left(n_x - \frac{1}{2}\right) S_x^1.
\end{aligned}
\end{equation}
The labels $H_0^f$, $H_{\rm int}^f$, and $H_{\rm int}^{fb}$ are applied as before, and the Heisenberg spin term is labeled $H_0^b + H_{\rm int}^b$, where $H_0^b = J\sum_{\left\langle xy\right\rangle} S_x^3 S_y^3$ as before, and $H_{\rm int}^b = J\sum_{\left\langle xy\right\rangle} \left(S_x^1 S_y^1 + S_x^2 S_y^2\right)$. For antiferromagnetism we require $J\geq 0$. While in the Ising interaction model every insertion of $S_x^1$ had to come in pairs on the same site $x$, there is no longer such a strong requirement for this model. We further restrict $h_x \geq 0$ for all $x$ (or $h_x \leq 0$ for all $x$) in this model to avoid sign problems.

We again split $H = H_0 + H_{\rm int}$, this time with $H_0 = H_0^f + H_0^b$, and $H_{\rm int} = H_{\rm int}^b + H_{\rm int}^f + H_{\rm int}^{fb}$. Again performing the unitary transformation in (\ref{transf}), we see that the antiferromagnetic interaction $H_0^b + H_{\rm int}^b$ is transformed in the following ways:
\begin{equation}
H_0^{U,b} = H_0^b =J\sum_{\left\langle xy \right\rangle} S_x^3 S_y^3, \qquad
H_{\rm int}^{U,b} =  - \frac{J}{2}
\sum_{\left\langle xy \right\rangle} \left( S_x^+ S_y^- + S_x^- S_y^+\right),
\label{spinsplit}
\end{equation}
and as we knew from before, $H_0^{U,f} = H_0^f$, $H_{\rm int}^{U,f} = H_{\rm int}^f$, and $H_{\rm int}^{fb}$ is given in (\ref{trans}). Expanding the partition function, as we did before, we get the following expressions:
\begin{equation}
Z = \sum_{m,\left\{b\right\}} \sum_{n,\left\{h\right\}} \sum_{k,\left\{z\right\}} \int \left[dt\right] 
\quad G_b\left[n,\left\{d\right\},k,\{z\}\right] G_f\left[k,\left\{z\right\}, m, \left\{b\right\}\right],
\label{pfmod2}
\end{equation}
where the fermionic pieces are the same as before, and spin pieces are now given by
\begin{equation}
G_b\left[n,\left\{d\right\},k,\{z\}\right] \ =\ (-1)^{k+n}{\rm Tr}\Big(... h_{z_1}S_{z_1}^1...H^{U,b}_{\rm int}(d_1)...h_{z_2}S_{z_2}^1 ... H^{U,b}_{\rm int}(d_n) ... h_{z_k}S_{z_k}^1 ...\Big).
\label{gsmod2}
\end{equation}
Now the trace depends on $n$ insertions of $H_{\rm int}^{U,b}(d\equiv \langle xy\rangle)$
$= -(J/2) \left( S_x^+ S_y^- + S_x^- S_y^+\right)$, which are nearest neighbor spin flips, and as before $k$ insertions of $h_zS_z^1$ with the free propagator $\mathrm{e}^{-t H_0^b}$ in between represented as ellipses. This configuration is labeled with $[n,\{d\},k,\{z\}]$. The fermionic trace is the same as before and is given by (\ref{ftrace}), where each configuration is labeled by $[k,\{z\},m,\{b\}]$. As in the previous example it is positive. The bosonic trace is also positive since $k$ turns out to be even and the $(-1)^n$ factor is canceled by the negative signs that appear in front of $H_{\rm int}^{U,b}(d)$. The trace in the spin space is evaluated by inserting a complete set of states in the $S^z$ basis as before. Each insertion of $S^1_z$ flips a single spin on the site $z$, while the insertion of $H_{\rm int}^{U,b}(d)$ flips both spins on the bond denoted by $d$. In the language of hardcore bosons, $S^3_z$ acts as either a creation or an annihilation event while $H_{\rm int}^{U,b}(d)$ acts as an event where the boson hops (see the right side of figure \ref{worldlines}). Since every creation event needs to be accompanied by an annihilation event, $k$ must be even. Thus, again there is no sign problem in the CT-INT expansion when spins are represented in the worldline representation.
\section{$\mathbb{Z}_2$ Gauge Theory}
We can extend these ideas to $\mathbb{Z}_2$ gauge theories as well. Consider
\begin{equation}
H = -t\left(\sum_{\left\langle xy\right\rangle} c_x^\dagger \sigma_{xy}^3 c_y + c_y^\dagger \sigma_{xy}^3 c_x \right) - h\sum_{\left\langle xy\right\rangle} \sigma_{xy}^1 + \sum_{\rm plaquettes} \sigma_a^3 \sigma_b^3 \sigma_c^3 \sigma_d^3.
\end{equation}
The first term, $H_0^{fb}$, is the covariant free part. We also add a gauge field term, $H_{\rm int}^b$, with fields purely in the $x$-direction. Finally, the last term, $H^p$, is a sum over plaquettes, where $a, b, c$ and $d$ label the bonds that make up each of the plaquettes.

It is straightforward to confirm that for the gauge transformation operator
\begin{equation}
G_x = \sigma_{x_1}^1 \sigma_{x_2}^1 \sigma_{x_3}^1 \sigma_{x_4}^1 \eta_x \left(-1\right)^{n_x},
\end{equation}
where $x_1, x_2, x_3$ and $x_4$ label the bonds that touch the point $x$, the Hamiltonian is invariant under gauge transformations, which entail $G_x^\dagger c_x G_x = -c_x$, $G_x^\dagger c^\dagger_x G_x = -c_x^\dagger$, $G_x^\dagger \sigma_{x_n}^1 G_x = \sigma_{x_n}$, and $G_x^\dagger \sigma_{x_n}^3 G_x = -\sigma_{x_n}^3$. The spinful version of this model was recently considered in \cite{Gazit16}.

We again use CT-INT, and to enforce Gauss's law, make use of the following operator:
\begin{equation}
P = \prod_x \frac{1}{2}\left(1 + \eta_x \left(-1\right)^{n_x} \sigma_{x_1}^1 \sigma_{x_2}^1 \sigma_{x_3}^1 \sigma_{x_4}^1 \right).
\end{equation}
Setting $H_0 = H_0^{fb} + H^p$, the partition function is given by
\begin{equation}
\begin{aligned}
Z &= \sum_{\left\{k\right\}} (-1)^k {\rm Tr}\left(P e^{-\left(\beta - t_1\right)H_0} H_{\rm int}^b ... H_{\rm int}^b e^{-t_k H_0}\right) \\
&= \sum_{\left\{k\right\},\left\{m_x\right\}}\left(\frac{1}{2}\right)^N (-1)^k {\rm Tr}\left(\prod_x\left(\eta_x\left(-1\right)^{n_x}\sigma_{x_1}^1 \sigma_{x_2}^1 \sigma_{x_3}^1 \sigma_{x_4}^1\right)^{m_x} e^{-\left(\beta - t_1\right)H_0} H_{\rm int}^b ... H_{\rm int}^b e^{-t_k H_0}\right) \\
&= \sum_{\left\{k\right\}, \left\{s\right\}, \left\{m_x\right\}} C\left(\left\{s\right\}\right) {\rm Tr_f}\left(\prod_{x}\left(\eta_x\left(-1\right)^{n_x}\right)^{m_x} e^{-\left(\beta - t_1\right)H_0^{f}\left(s_1\right)} ... e^{-t_k H_0^{f}\left(s_k\right) }\right).
\end{aligned}
\label{partz2}
\end{equation}
Here $\left\{m_x\right\}$ is a set of $N$ variables, where $m_x \in \left\{0,1\right\}$ for every site $x$. There must always be an even number of $m_x = 1$ values to get a nonzero value in the fermionic sector. Because of this, we know that there must be an even number of $H_{\rm int}^b$ insertions, and so $k$ is even and the spin sector will give a positive contribution. We cannot factorize into two independent traces this time, so the last line of (\ref{partz2}) shows the results after the spin portion has again been evaluated in the spin-$z$ basis, leaving only fermionic operators. All constant factors, including the $\left(-1\right)^k$ and the $\left(1/2\right)^N$, are absorbed into the $C\left(\left\{s\right\}\right)$ prefactors, which we have argued must be positive.

For the fermionic operators, we note that
\begin{equation}
\left(-1\right)^{n_x} = -2\eta_x \left(n_x - \frac{1}{2}\right),
\end{equation}
and so we see that the partition function is
\begin{equation}
Z = \sum_{\left\{k\right\}, \left\{s\right\}} C\left(\left\{s\right\}\right) {\rm Tr_f}\left(\prod_{x}\left(-2\eta_x \left(n_x - \frac{1}{2}\right)\right)^{m_x} e^{-\left(\beta - t_1\right)H_0^{f}\left(s_1\right)} ... e^{-t_k H_0^{f}\left(s_k\right) }\right).
\end{equation}
Noting that $H_0^f\left(s\right) = -t\sum_{\left\langle xy\right\rangle} \lambda_s\left(c_x^\dagger c_y + c_y^\dagger c_x\right)$, we find that this model now simply has staggered chemical potential terms coming from the Gauss's law projection operator and thus falls under the purview of \cite{Wang25,Wei:2016sgb,Li:2016gte}. There is thus no sign problem in this expansion.

\section{Conclusions}
In this work we have explored several ways in which the CT-INT formulation in the Hamiltonian approach reveals more sign-problem solutions in lattice field theory, including models with interacting spins and fermions as well as lattice gauge theories. While we have given a couple of examples, many more models are solvable with these techniques, several of which are given in \cite{Huffman:2016nrx}.

\section*{Acknowledgements}
The material presented here is based upon work supported by the U.S. Department of Energy, Office of Science, Nuclear Physics program under Award Number DE-FG02-05ER41368, and is also supported by a National Physical Science Consortium fellowship.


\begin{thebibliography}{99}
\bibitem{Huf14}
E.~F.~Huffman and S.~Chandrasekharan, Phys.\ Rev.\ B 89 (2014) 111101.

\bibitem{Li:2014tla}
Z.~X.~Li, Y.~F.~Jiang and H.~Yao, Phys.\ Rev.\ B 91 (2015) 241117.

\bibitem{Wang25}
L.~Wang, Y.~H.~Liu, M.~Iazzi, M.~Troyer, and G.~Harcos, Phys.\ Rev.\ Lett.\ 115 (2015) 250601.

\bibitem{Wei:2016sgb} 
  Z.~C.~Wei, C.~Wu, Y.~Li, S.~Zhang and T.~Xiang,
  Phys.\ Rev.\ Lett.\  116, (2016) 250601.

\bibitem{Li:2016gte} 
  Z.~X.~Li, Y.~F.~Jiang and H.~Yao,
  arXiv:1601.05780 [cond-mat.str-el] (2016).
	
	\bibitem{Wang:2015rga} 
  L.~Wang, M.~Iazzi, P.~Corboz and M.~Troyer,
  Phys.\ Rev.\ B 91, (2015) 235151.
	
	
\bibitem{Chandrasekharan:2012fk} 
  S.~Chandrasekharan,
  Phys.\ Rev.\ D {\bf 86}, (2012) 021701.
	
	\bibitem{Huffman:2016nrx} 
  E.~Huffman and S.~Chandrasekharan,
  Phys.\ Rev.\ E 94, (2016) 043311.
	
	\bibitem{Wang:2014cbw} 
  L.~Wang, P.~Corboz and M.~Troyer,
  New J.\ Phys.\ 16, (2014) 103008.

\bibitem{Li:2014aoa} 
  Z.~X.~Li, Y.~F.~Jiang and H.~Yao,
  New J.\ Phys.\  17, (2015) 085003.
	
	\bibitem{Gazit16}
S.~Gazit, M.~Randeria, A.~Vishwanath, arxiv:1607.03892 [cond-mat.str-el] (2016).
	





\end{thebibliography}
\end{document}